\documentclass[
aps,
prl,
groupedaddress,
superscriptaddress,
floatfix,
twocolumn,
notitlepage
]{revtex4-1}

\usepackage{graphicx}% Include figure files
\usepackage{amsmath,amssymb}
\usepackage{braket}
\usepackage{hyperref} % hypertext capabilities
\usepackage{subfigure}
\usepackage{bm} % bold math
\usepackage{mathtools} % dcases (uncompressed fractions)
\usepackage{float}

\newif\iffigures
\figurestrue

\iffigures
\usepackage{cprotect}
\fi

\usepackage{color}
\newcommand{\drg}[1]{\color{black}{#1}} % Dmitry Gulevich commentaries

\begin{document}
	
%\preprint{}

\title{Bridging the Terahertz gap for chaotic sources with superconducting junctions}
	
\author{D. R. Gulevich} 
\affiliation{ITMO University, St. Petersburg 197101, Russia}

\author{V. P. Koshelets}
\affiliation{Kotel’nikov Institute of Radio Engineering and Electronics, Russian Academy of Science, Moscow, 125009, Russia}
	
\author{F.\,V.~Kusmartsev}
\affiliation{Department of Physics, Loughborough University, United Kingdom}
\affiliation{ITMO University, St. Petersburg 197101, Russia}

\date{\today}

\begin{abstract} 
We observe a broadband chaotic signal of Terahertz frequency emitted from a superconducting junction. 
The generated radiation has a wide spectrum reaching 0.7 THz and power sufficient to drive on-chip circuit elements. 
To our knowledge, this is the first experimental observation of a high-frequency chaotic signal emitted by a superconducting system which lies inside the Terahertz gap.
{\drg
Our experimental finding is fully confirmed by the numerical modeling based on the microscopic theory and reveals the unrealized potential of superconducting systems in chaos-based Terahertz communication, fast generation of true random numbers and non-invasive Terahertz spectroscopy applicable to physical, chemical and biological systems.
}
\end{abstract}

%Our finding reveals the unrealized potential of superconducting systems for various chaos applications 

%\maketitle
{\let\newpage\relax\maketitle}

%-----------------------------------------------------------------------------------------------
%-----------------------------------------------------------------------------------------------
%{\it Introduction.}
%-----------------------------------------------------------------------------------------------
%-----------------------------------------------------------------------------------------------
Since the call for bridging the Teraherz gap 
had rocketed~\cite{Ferguson-2002, Siegel-2002},
many 
%physical 
systems were proposed
%~\cite{Kohler-2002,Williams-NP-2007,Hubers-2005, Chas-2009,Lus-2005,Li-2010,gunn-diodes,freq-mult,Jain-1984, phase-locking, Booi-1996, on-chip, Darula-1999, Barbara-1999, Koshelets-Shitov, Kuzmin-2004, Ozyuzer-2007, Wang-PRL-2010, Kashiwagi-2012, Welp-2013, Koshelets-spectrometers, array-1} 
as sources of Terahertz radiation (from 0.3 to 3 THz).
%in the Terahertz range (from 0.3 to 3 THz).
The years of intense research
% in the Terahertz field
yielded quantum cascade lasers~\cite{Kohler-2002,Williams-NP-2007}, 
semiconductor lasers and transistors~\cite{Hubers-2005, Chas-2009, Lus-2005, Li-2010}, Gunn diodes~\cite{gunn-diodes},
frequency multipliers~\cite{freq-mult}, generators based on low
and high temperature superconductors~\cite{Jain-1984, phase-locking, Booi-1996, on-chip, Darula-1999, Barbara-1999, Koshelets-Shitov, Kuzmin-2004, Ozyuzer-2007, Wang-PRL-2010, Kashiwagi-2012, Welp-2013, Koshelets-spectrometers, array-1}.
With the Terahertz gap for radiation sources gradually filling up with new practical and compact devices, there is yet one more Terahertz gap to fill: the one for {\it chaotic} radiation sources.
The need for them
%chaotic sources in Terahertz frequency range 
arises in cryptography which uses chaos to encrypt information~\cite{Hayes-1993, Argyris-2005, Ren-2013},
high-resolution THz spectroscopy~\cite{Sob-2012,Koshelets-THz-spectroscopy,Vaks-2017,Sun-2017} and computer engineering which demands high-rate true random number generators~\cite{Fisher-2012}.
Although alternative technologies for generation of true random numbers exist, which rely on quantum effects~\cite{Qi-2010, Symul-2011, Ivanova-2017} and various entropy sources~\cite{Gabriel-2010, mobile-phone, Danger-2009}, the chaos-based systems 
%generally lead the field, being 
enable to achieve higher generation rates~\cite{Uchida-NP-2008, Reidler-PRL-2009, Kanter-NP-2010, Sci-2015}.

Chaotic solutions
%of the driven 
%the sine-Gordon model describing 
in theoretical models of small Josephson junction (SJJ) 
%under the rf radiation 
were first found theoretically in Ref.~\cite{Huberman-1980}.
The early studies~\cite{Pedersen-1981, Gubankov-chaos, Kautz-1985} were primarily centered to rf driven SJJ, while the more recent ones~\cite{Botha-2013, Shuk-2015, Botha-2015, Pank-2017} focused on dynamics of a finite number of coupled oscillators. 
Despite a progress in achieving a chaotic state of extended Josephson systems~\cite{Ustinov-chaos},
%an extended Josephson system has been achieved 
%In the past, 
all existing experimental studies of chaos in superconducting systems so far dealt with microwave frequencies 
%limited by 
($<$ 100 GHz). %which 
This is significantly below the interest of modern Terahertz science.

In this Letter, we present our 
%experimental and theoretical 
results for generation of high-frequency broadband chaos reaching 0.7 THz in frequency.
A highly chaotic regime is achieved in a T-shaped Josephson junction which we here refer to as 
T-junction flux-flow oscillator (TFFO), see Fig.~\ref{fig:tffo}.
In contrast to the conventional flux-flow oscillator (FFO)~\cite{Nagatsuma1}
consisting of a tapered Josephson transmission line~\cite{FFO-tapered}, 
%of an elongated rectangular shape (usually, tapered at the ends~\cite{FFO-tapered}), 
TFFO is made of 
two perpendicular lines coupled via a T-junction: the main (MJTL) and additional (AJTL) Josephson transmission lines.
While it has been understood in the past that the T-junction brings new effects to the dynamics of a single fluxon by slicing it into pieces~\cite{flux-cloning, Gulevich-2D, Gulevich-NJP, DGulevich-pump}, introducing T-junction in a standard FFO yields a completely unexpected result: it transforms one of the narrowest linewidth source
%in the Terahertz frequency range 
ever created~\cite{Koshelets-Shitov} to the opposite extreme -- a highly chaotic generator with very broad radiation spectrum.
Indeed, results that we present in this Letter demonstrate the onset of chaos in almost all of the current-voltage characteristics (IVC) of the device which includes the region of Terahertz gap 0.3-0.9 THz 
%inside the Terahertz gap
where compact and practical devices are highly wanted.

%from a different perspective.
%the TFFO, however brings new surprise:
%Incidentally, while the conventional FFO is known to be the narrowest linewidth radiation source in the Terahertz frequency range~\cite{Koshelets-Shitov}, introduction of a T-junction transforms it 
%to the opposite extreme of a chaotic generator with a very broad radiation spectrum:
%our results presented below demonstrate the onset of chaos in almost all of the flux-flow current-voltage characteristics (IVC) including the region 0.3-0.9 THz inside the Terahertz gap
%where compact and practical devices are highly desirable.

%Owing to the nontrivial two-dimensional layout of such system, 
%It has been predicted in the past that the lateral dynamics of the superconducting phase inside a T-junction is very nontrivial owning to the flux cloning~\cite{flux-cloning, Gulevich-2D}:
%when a fluxon passes through a T-junction, it branches off another fluxon in the perpendicular direction.
%becomes crucially important 
%and results in a very peculiar motion of fluxons in its two perpendicular directions
%coined flux cloning~\cite{flux-cloning, Gulevich-2D}.
%When a fluxon in the main Josephson transmission line (MJTL, see Fig.~\ref{fig:tffo}) is
%passing through a T-junction, it branches off another fluxon in the additional Josephson transmission line (AJTL): a phenomenon coined flux cloning~\cite{flux-cloning, Gulevich-2D}.

{\drg
High-frequency chaos generators promise to impact several independent areas of research and technologies.
%With frequency of chaos generation entering the Terahertz gap by means of the TFFO proposed in this Letter, the science of chaotic systems promises to impact several independent areas of research and technologies.
%Apart from the purely practical aspects, 
%The high frequencies of electromagnetic waves in the THz region enable high data rates as compared to the radio waves, %whereas several transparency windows in the Earth atmosphere fall in the THz domain of the electromagnetic spectrum. This makes compact THz chaos generators particularly suitable for application in 
%will make viable chaos-based communications.
The high data rates of THz-frequency chaotic signals 
% compared to the radio waves, %whereas several transparency windows in the Earth atmosphere fall in the THz domain of the electromagnetic spectrum. This makes compact THz chaos generators particularly suitable for application in 
will make viable chaos-based communications.
%, apart from various applications of a true random generator and and the novel type of chaos-based spectroscopy. 
Development of THz chaotic sources will open opportunities for the chaos-based molecular and biological spectroscopy where
sources of high-frequency broadband noise are needed~\cite{Koshelets-THz-spectroscopy}.
% a chaotic system is used to produce a useful THz noise~\cite{Koshelets-THz-spectroscopy}. 
The resulting chaos-based THz spectroscopy can be used to probe physical and chemical processes in biological and living systems which otherwise would be impossible or extremely difficult to observe using the tools of the conventional spectroscopy: transient biological structures, unstable molecules and chemical reactions. 

TFFO is very different from all existing superconducting systems. %studied previously. Most remarkably, while it has been always given as granted that 
%If the characteristic frequency of a Josephson junction is the Josephson frequency (which varies with the applied voltage $V$ and equals $f_V=2eV/h$),
Most notably,
the electric current oscillations in TFFO do not occur at the Josephson frequency ($f_V=2eV/h$, which varies with the applied voltage~$V$). Moreover, %as our results below show, 
in some regimes,
the TFFO frequency spectrum 
of Josephson current oscillations does not even peak at the Josephson frequency.
%All these 
%That creates tremendous difficulties to describe the phenomenon %: %in the theoretical description: 
%one needs to take into account the 
%where there is 
%The effect arises due to a coupling between the tunnel currents and electromagnetic waves at all frequencies. 
Because the coupling between the tunnel currents and electromagnetic waves is not at a single frequency, but 
involves all possible modes of the whole spectrum,
%occurs have no dominant frequency,
%of all frequencies of the spectrum,
% at all frequencies of the spectrum, 
this creates tremendous theoretical difficulties which can not be resolved with the standard approaches~\cite{Pankratov-2007, FFO-MCQTN}. 
%and impossible %the phenomenon and not the single one as 
%in other approaches% the existing models
%~\cite{Pankratov-2007, FFO-MCQTN}. 
%We find a work around this by using the full microscopic description of the junction~\cite{Werthamer}.
}

%From the fully fundamental point of view, TFFO stands out as a continuous system with an infinite degrees of freedom which undergoes the Ruelle-Takens-Newhouse transition to chaos~\cite{Eckmann}. All these make TFFO a remarkable system both from a very practical as well as completely fundamental aspects.

\iffigures
\begin{figure}[t!]
	\begin{center}
\includegraphics[width=3.4in]{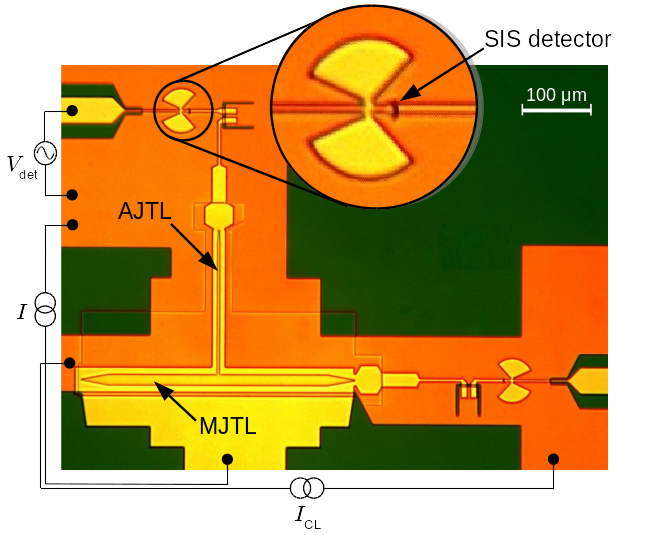}
		\caption{\label{fig:tffo} 
Optical microphotograph of the experimental sample of T-junction flux-flow oscillator (TFFO) fabricated using the Nb-AlN-NbN technology~\cite{NbN-Dmitriev, Nb-NbN-FFOs}. The TFFO consists of the additional Josephson transmission line (AJTL) coupled to the main Josephson transmission line (MJTL) in the form of a T-junction. 
{\drg TFFO is coupled to two SIS detectors, one at the AJTL and the other at the MJTL ends.}
Zoom of the SIS detector coupled to the AJTL is shown on the inset. 
%(detector 1) and MJTL (detector 2) are shown on the insets. 
%The AJTL and MJTL are tapered at the ends as in the routinely employed FFO systems~\cite{FFO-tapered}. 
%which was used to have a beneficial effect on the performance of a FFO and facilitates its matching to SIS detector.
		}		
	\end{center}		
\end{figure}
\fi

\iffigures
\begin{figure*}
	\begin{center}
\includegraphics[width=7.0in]{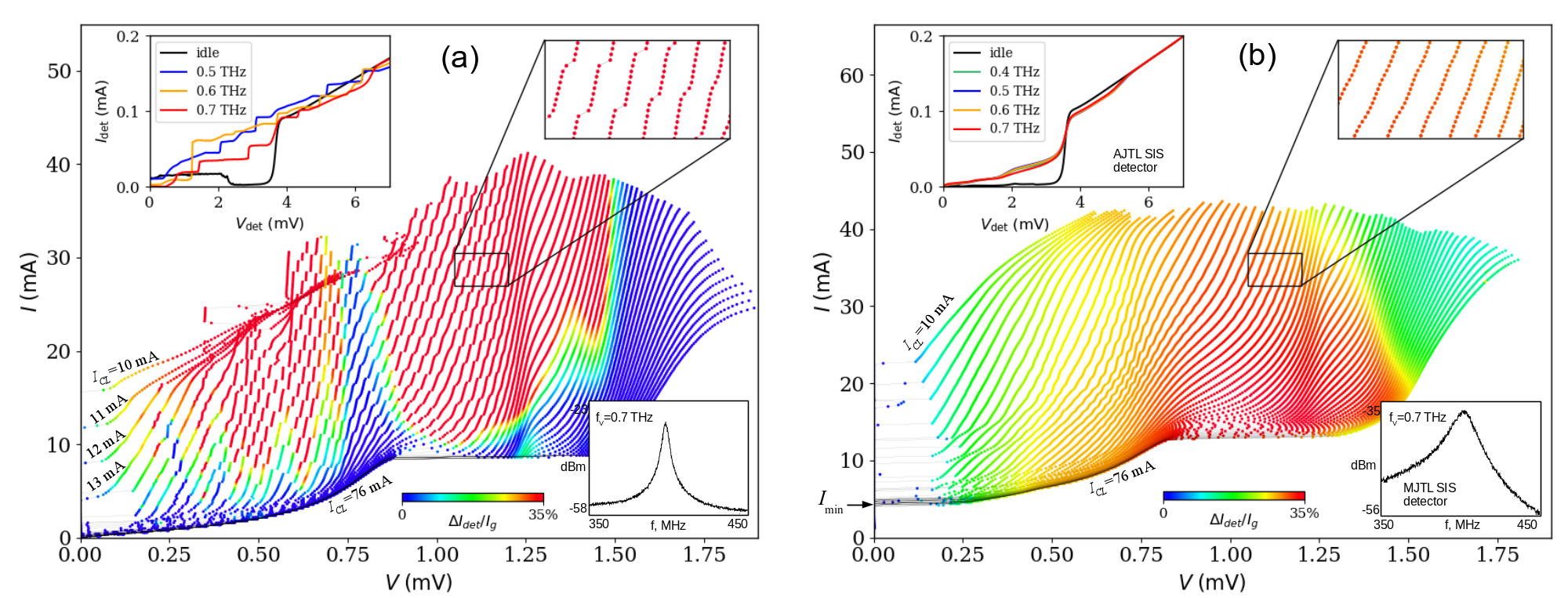}
		\cprotect\caption{\label{fig:ivc-exp} 
%		Experimental results for Nb-AlN-NbN junctions. 
(a)	Current-voltage characteristics (IVC) of the control sample -- FFO of a standard layout. A set of IVC curves is measured at different values of the control line current $I_{\rm CL}$ ranging from 10 mA (the leftmost curve) to 76 mA (the rightmost curve) with an increment 1 mA. The color scale corresponds to the rise $\Delta I_{\rm det}$ in the dc current through the SIS detector from 0 to 35\% of the full step $I_g$ at the gap voltage (the precise definition of $I_g$ is given in Ref.~\cite{Ermakov-2001}). The data where dc current rises above the 35\% threshold is painted by the same (red) color as the 35\% rise. Note that Fiske steps are pronounced up to the boundary voltage $V_g/3$ (about 1.2 mV). Response of the SIS detector to the radiation emitted by the FFO is shown on the inset.  The black line corresponds to an autonomous operation of the SIS detector, the blue, orange and red lines represent pumping by the FFO at voltage $V$ specified on the legend by the Josephson frequency $f_V=2eV/h$. Note the presence of sharp Shapiro steps in the IVC of the detector 
and quasiparticle steps whose positions correspond exactly to the Josephson frequency of FFO.		
(b) IVC of one of the fabricated experimental samples of TFFO. 
As compared to the standard FFO, the Fiske steps are practically absent while the IVC curves are very smooth even below $V_g/3$. 
The inset shows IVC of the SIS detector coupled to the AJTL end taken at different values of TFFO voltage. Note that the SIS detector is driven not on the Josephson frequency but instead exhibits very broad steps centered at approximately the same voltage. {\drg The inset on the bottom right of the panel (b) shows the line profile 
%in the region of high voltages and high magnetic fields 
measured from the MJTL at the Josephson frequency $f_V=0.7\rm\;THz$. The red and blue lines correspond to the free-running and phase-locked regimes, respectively. This is about the narrowest radiation line we could measure from the TFFO. The linewidth increases rapidly with decreasing the frequency below 0.7 THz. 
No detectable line was measured for radiation from AJTL at all, but a featureless flat noise level. For comparison, the line profile for the standard FFO is shown on the inset at the right of the panel~(a).}
		}		
	\end{center}		
\end{figure*}
\fi

%-----------------------------------------------------------------------------------------------
%-----------------------------------------------------------------------------------------------
%{\it Experimental results.}
%-----------------------------------------------------------------------------------------------
%-----------------------------------------------------------------------------------------------
Several samples of TFFO (one of them presented in Fig.~\ref{fig:tffo})
were fabricated using the Nb-AlN-NbN technology~\cite{NbN-Dmitriev, Nb-NbN-FFOs} which allows to achieve high radiation frequencies~\cite{footnote1}.
%, as compared to Nb-$\rm AlO_x$-Nb junctions~\cite{footnote1}.
A control sample of the standard FFO with length equal to the length of MJTL in TFFO ($400\rm\;\mu m$) has been fabricated on the same chip for testing the setup.

%We will first discuss operation of the control sample, see Fig.~\ref{fig:ivc-exp}a.
The IVC of a standard FFO consists of a set of curves which correspond to a fixed value of the electric current 
% in the dedicated 
in a control line $I_{\rm CL}$ used to induce the external magnetic field.
% inside the sample.
In the region of moderate voltages (up to 1.2 mV in Fig.~\ref{fig:ivc-exp}a) FFO exhibits a series of Fiske steps (FS) arising %due to the resonances with standing electromagnetic modes. 
when the standing electromagnetic waves in the junction resonate with the rate at which fluxons enter and leave the junction.
% (the Josephson frequency $f_V$). 
Typically, for junctions of moderate lengths in the range 200-600 $\mu\rm m$, FS are pronounced up to the voltage $V_g/3$~\cite{sc-Koshelets, footnote2} ($V_g$ is the gap voltage) where an abrupt increase of the damping suppresses propagation of electromagnetic modes.
%($V_g$ is the gap voltage defined by the sum of two superconducting energy gaps $(\Delta_1+\Delta_2)/e$). 
It occurs due to the phenomenon of self-coupling  
 when the tunneling of quasiparticles through the barrier
% between the two superconductors 
 is enhanced by absorption of photons generated by the junction~\cite{Werthamer}.
% photons generated inside the junction~\cite{Werthamer}.
The radiation emitted from the FFO is detected by a superconductor-insulator-superconductor junction (SIS).
% coupled to the FFO end.
The IVC of SIS irradiated by the FFO exhibits
a series of sharp and prominent Shapiro steps and quasiparticle steps
which correspond exactly to the Josephson frequency at which the FFO is driven. 
{\drg
The SIS can also be operated as a mixer by mixing the signals from the FFO and from a local oscillator.
% (see, e.g. the linewidth measurement on the inset to Fig.~\ref{fig:ivc-exp}a).
} 
%The described operation of the control sample is in full agreement with the previous studies of FFO made with Nb-AlN-NbN technology~\cite{Nb-NbN-FFOs}.

%The resulting down-converted spectrum can be analyzed 1 GHz bandwidth at a time.
%allows to measure the radiation spectrum of the system when the radiated signal i and down-converted to low frequencies .
%The down-converted spectrum is measured with 1 GHz bandwidth at a time.

In contrast,
% to the IVC of a standard FFO, 
the IVC of TFFO is drastically different, see Fig.~\ref{fig:ivc-exp}b. 
A striking feature is a non-zero return current $I_{\rm min}$ 
%at large values of 
%the control current $I_{\rm CL}$, 
which can be interpreted as a presence of a finite barrier for fluxons passing through the T-junction.
% The rest of the IVC is even more puzzling: 
 Moreover, all curves exhibit a remarkable suppression of FS and appear smooth even below the voltage $V_g/3$ where the standard FFO would exhibit pronounced steps (cf. Fig.~\ref{fig:ivc-exp}a). The observed suppression of FS
%  in IVC of TFFO 
  is understood as follows. 
In general, presence of a step on the IVC depends on whether a standing electromagnetic wave may form, i.e. if the traveling electromagnetic wave can propagate from one end to the other without significantly loosing its energy, and, whether there is a regularity in fluxon dynamics, i.e. fluxons enter and leave the junction at regular time intervals. 
In our system, while the first condition is satisfied as suggested by the study of the control FFO sample of the same length, 
it is the second one being violated and signifies a very irregular fluxon dynamics.
The noted irregularity of fluxon dynamics %based on the shape of IVC curves of TFFO 
is further confirmed by the measurements of TFFO radiation using the two SIS detectors (see Fig.~\ref{fig:ivc-exp}).

IVCs of SIS detectors operated in the power detection mode
%each operated in the power detection and mixer modes. 
%coupled to the AJTL end. 
reveal very smooth quasiparticle steps and suggest broad radiation spectrum. The shapes of the quasiparticle steps of AJTL SIS detector (the upper inset to Fig.~\ref{fig:ivc-exp}b) suggest that the spectrum radiated from AJTL is at least 0.1 THz as broad.
Even more surprisingly, quasiparticle steps on the IVC of AJTL SIS are independent of the Josephson frequency $f_V$ (see also our results for other TFFO samples in Ref.~\cite{SM}).

%to center around a fixed frequency rather than the Josephson frequency $f_V$ at which the TFFO is driven.
%Such behavior can be explained by a very flat power spectrum of radiation from the AJTL, so that the frequency spectrum that reaches the SIS detector is fully determined by the properties of the matching circuitry rather than the TFFO voltage.

%Some samples of TFFO were fabricate to allow a direct measurement of the spectrum of radiation via SIS in the harmonic mixing regime. 

%To fully study the properties of the radiation emitted from AJTL, two kind of samples were fabricated, where SIS junctions were operating as 

%To perform a direct measurement of the spectrum of TFFO radiation we some samples of TFFOs were coupled to SIS junctions operating in the harmonic mixer regime instead of power detection regime shown above. 

{\drg
The spectrum radiated from the MJTL and AJTL ends (see Fig.~\ref{fig:tffo}) was analyzed by both SIS operated %the SIS junctions 
in the mixer mode.
Although, the maximal bandwidth of the down-converted spectrum was limited from above by 1~GHz, we %were able to
managed to scan this frequency window over the whole frequency range.
% of interest. 
%While a hint of 
A broad line has been %could be 
identified in the spectrum of MJTL radiation in high voltage and high magnetic field region of the IVC (see the inset to Fig.~\ref{fig:ivc-exp}b): this was the narrowest line we could detect from TFFO as its linewidth increases rapidly with decreasing the applied voltage. No detectable peak at all was observed in the spectrum of AJTL radiation, rather, a featureless flat noise level within the 1 GHz bandwidth.
This agrees with our measurement of SIS detectors operated in the power detection regime presented above.
%where very broad steps were observed on their IVCs, while the steps of the IVC of AJTL SIS were independent of the Josephson frequency.
% (see the inset on Fig.~\ref{fig:ivc-exp}b and Ref.~\cite{SM}).
%An example of the line at $f_V=0.7\rm\;THz$ measured at the MJTL is shown on the inset in Fig.~\ref{fig:ivc-exp}b. This is the narrowest line we could detect from TFFO as its linewidth increases rapidly with decreasing the applied voltage, while no line was detected from AJTL at all.
}

%The radiation spectra obtained by operating SIS junctions in the mixer mode 
%These spectra are in agreement with those obtained from  of the SIS junctions.
%Results of these measurement are in agreement with the conclusions based on IVC of SIS junctions.
%of radiation from TFFO by a SIS junction operating as a mixer. 
%While the measured down-converted spectrum radiated from MJTL was found to be very wide,
% and significantly larger than that of the standard FFO, .
% where the radiation spectrum measured at $f_V=0.7\rm\;THz$ is shown. 

%The experimental measurements 
Our measurements of the TFFO radiation by both SIS detectors operated in the power detection and mixing modes, supported by the absence of FS, indicate a very chaotic nature of the emitted radiation.
% emitted from  TFFO. 
%The absence of FS on the IVC of TFFO supported by the detailed analysis of 
We have found that the chaotic spectrum from AJTL persists 
%at every point of the flux-flow 
on all the IVC, all the way up to the high-frequency region near the Nb superconducting gap frequency (about 0.7 THz)
% and beyond, 
 where the onset of damping hinders operation of the device. 
%Below we present a theoretical model of TFFO which agrees with our experimental findings and helps to shed light on the mechanisms of chaos generation.

\iffigures
\begin{figure}
	\begin{center}
\includegraphics[width=3.3in]{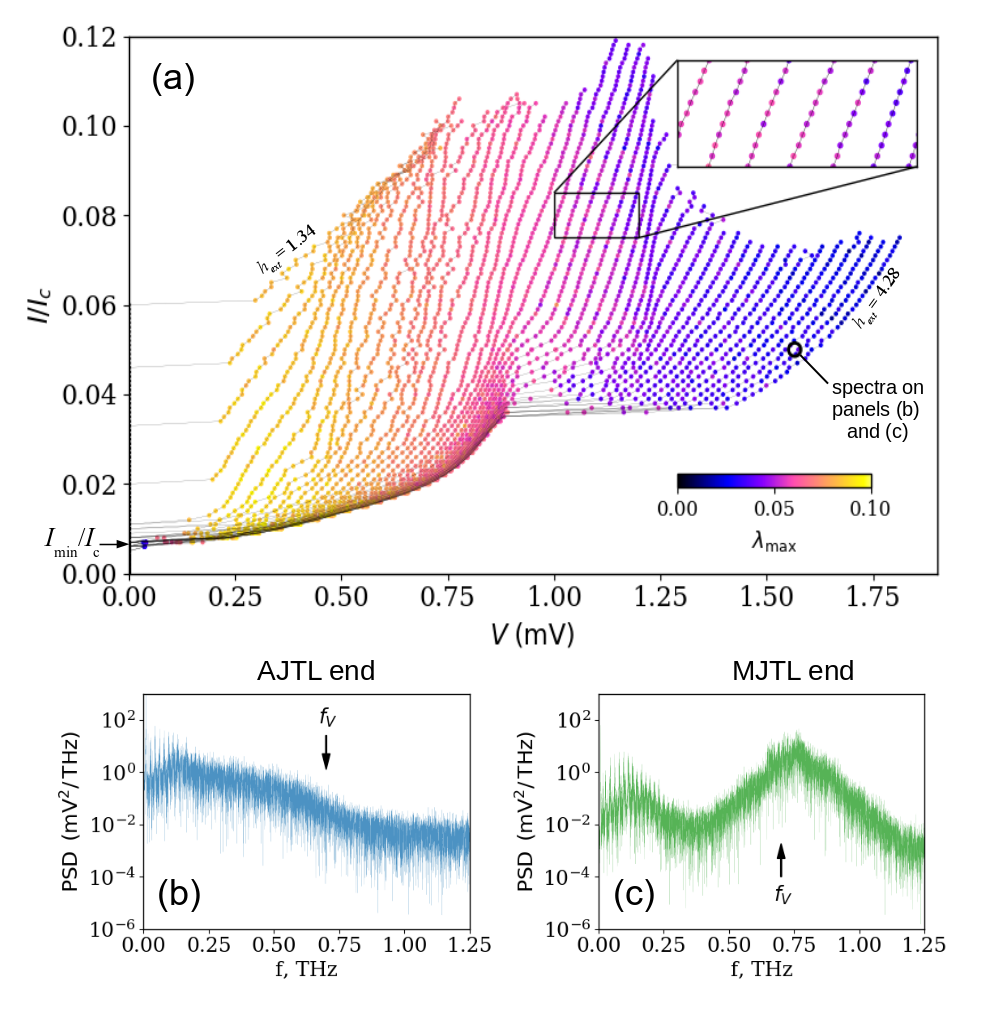}
		\cprotect\caption{\label{fig:ivc-th} 
(a) Theoretical current-voltage characteristics (IVC) of T-junction flux-flow oscillator (TFFO) calculated using the finite element library \verb|deal.II|~\cite{dealii-original} in conjunction with the microscopic tunneling library \verb|MiTMoJCo|~\cite{MiTMoJCo-CPC,mitmojco}.
The curves are obtained at different values of the external magnetic field $h_{\rm ext}$ 
changing from 1.34 to 4.28 with an increment 0.07 in normalized units (see Ref.~\cite{SM} for details). 
The color scale represents the numerically calculated maximal Lyapunov exponent $\lambda_{\rm max}$ 
which indicates the presence of chaotic dynamics at every point of the flux-flow IVC.
Panels (b) and (c) display the chaotic spectra at the AJTL and MJTL ends at the Josephson frequency $f_V=0.7\rm\;THz$. The corresponding point on the IVC is marked by a black circle in (a).
		}		
	\end{center}		
\end{figure}
\fi

%-----------------------------------------------------------------------------------------------
%-----------------------------------------------------------------------------------------------
%{\it Theoretical results.}
%-----------------------------------------------------------------------------------------------
%-----------------------------------------------------------------------------------------------
%The major obstacle in the theoretical description of TFFO arises from 
%the coupling between quasiparticle tunnel currents and electromagnetic waves generated inside the junction, that is, the 
%self-coupling.
Standard theoretical models of long Josephson junctions based on the 
the perturbative sine-Gordon equation take into account the self-coupling 
as a phenomenological modification of the damping parameter assuming the spectrum is dominated by a single harmonics at the Josephson frequency~\cite{Pankratov-2007, FFO-MCQTN}.
It is evident that while such approach is justified in case of a standard FFO exhibiting a narrow radiation linewidth,
it would a priori fail in the description of TFFO because of a large number of competing electromagnetic modes as suggested by the broad spectrum observed in our experiment.
%al measurement of IVC of the SIS junction.
%experimentally observed broad radiation spectrum.
In this case, coupling between the generated radiation and tunnel currents 
occurs for many excited modes of the chaotic spectrum simultaneously.
% because no single frequency dominates the others. 
We term such dynamical regime
the {\it chaotic self-coupling} in contrast to the conventional self-coupling
when only coupling to the single electromagnetic mode at $f_V$
%the Josephson frequency 
is relevant.

To obtain the theoretical treatment of TFFO in the regime of the chaotic self-coupling it is, therefore, 
essential to account for coupling between the quasiparticle tunnel currents and electromagnetic field oscillations of {\it all} the involved frequencies.
% rather than a single one as in the standard FFO models~\cite{Pankratov-2007, FFO-MCQTN}.
Such description is natural within the microscopic tunneling theory (MTT) of Josephson junction~\cite{Werthamer}.
Application of MTT to large Josephson junction has been recently pushed forward by the authors in Ref.~\cite{Gulevich-MTT,Gulevich-JLTP} and motivated the development of \verb|MiTMoJCo| code~\cite{MiTMoJCo-CPC,mitmojco}.
We used \verb|deal.II| finite element library~\cite{dealii-original,dealii-8.5} 
in conjunction with \verb|MiTMoJCo| 
to solve the integro-differential equation describing the two-dimensional model of TFFO.
The two-dimensional mesh was generated using~\verb|Gmsh| mesh generator~\cite{gmsh}. 
For evaluation of the tunnel currents we used 
smoothed tunnel current amplitudes calculated from the Bardeen-Cooper-Schrieffer 
theory for Nb-AlN-NbN junction made of superconductors with different energy gaps: Nb (1.4 meV) and NbN (2.3 meV) at $T=4.2$ K (see Ref.~\cite{SM} for further details).

Results of our numerical calculations are presented in Fig.~\ref{fig:ivc-th}. 
The IVC curves were obtained at different values of the parameter $h_{\rm ext}$
which describes the external magnetic field induced by the control line current $I_{\rm CL}$ in our experimental setup.
The numerical IVC exhibits the same features as the experimental IVC of TFFO: a non-zero return current $I_{\rm min}$
and the absence of FS.
We have checked that the suppression of the FS is not caused by the damping by running simulation for the standard FFO of the same length 
and observing well pronounced FS below $V_g/3$.
% in agreement with our experiments.
To reliably conclude about the chaotic regime we calculated the maximal Lyapunov exponent $\lambda_{\rm max}$ at every point of the numerical IVC of TFFO using the procedure of Benettin et al.~\cite{Benettin} generalized to the presence of memory~\cite{Benettin-memory}.
Results of our calculation are presented Fig.~\ref{fig:ivc-th}a.
% by the color scale.
As seen from the Figure, maximal Lyapunov exponent $\lambda_{\rm max}$ 
takes positive values 
in the whole IVC apart from the zero-voltage state.
% at~$V=0$.
%
The numerically calculated power spectra of the TFFO radiation
%normalized ac voltage at the AJTL and MJTL edges of TFFO 
are shown in Figs.~\ref{fig:ivc-th}b and c. 
In full agreement with our experiments, the spectrum of radiation from MJTL, while being very broad, shows some dependence on the Josephson frequency $f_V$, whereas the spectrum emitted from AJTL does not even peak at the Josephson frequency.
%Such behavior is in agreement with our experimental results where we observe no steps on Josephson frequency in the IVC of SIS detector coupled to AJTL (cf. the inset of Fig.~\ref{fig:ivc-exp}b).

%Indeed, in our experiments we see no detectable peak in the spectra from AJTL with the 1 GHz bandwidth, neither at the Josephson frequency nor other frequencies. To the contrary, as we have shown in Fig.~\ref{fig:ivc-exp}, the hint of a broad line could only be detected when operating the MJTL SIS junction as a mixer in high voltage and high magnetic field region. 

%However this is still about 2 orders of magnitude broader than the linewidth of the standard FFO.

To analyze the transition to chaos in our numerical simulations we introduced a control parameter $h_X$ which can be interpreted as a magnetic field component applied along the horizontal axis in Fig.~\ref{fig:tffo}. When $h_X=0$,
%no horizontal component is present, 
the TFFO exhibits a chaotic regime as has been demonstrated above. 
%However, 
The gradual increase of $h_X$
forces the system to enter a regime of periodic oscillations at $h_X\sim h_{\rm ext}$. This signifies that it is the dilute mixture of weakly bound fluxons 
%dwelling 
in AJTL which is responsible for the observed 
chaos.
%chaotic dynamics. 
Indeed, as long as fluxons in AJTL are arranged into a dense chain by
at high $h_X$,
%high values of $h_X$,
%a finite magnetic field $h_X$,  
regular oscillations are observed, whereas 
when fluxons become 
diluted at smaller $h_X$
%less influenced by the magnetic field $h_X$
the chaotic dynamics gradually builds up.
Our analysis shows that with decreasing $h_X$ the system undergoes the Ruelle-Takens-Newhouse route to chaos~\cite{Eckmann} by sequential torus splitting from a limit circle in the phase space to a 2D- and then 3D-torus which precede the transition to chaos.
%Interestingly, TFFO is thus one more physical realization of a system which undergoes Ruelle-Takens-Newhouse transition in violation to the classical Landau scenario~\cite{Landau}.

%-----------------------------------------------------------------------------------------------
%-----------------------------------------------------------------------------------------------
%{\it Conclusion.}
%-----------------------------------------------------------------------------------------------
%-----------------------------------------------------------------------------------------------
%We have demonstrated experimentally and confirmed theoretically generation of chaotic Terahertz-range signal from a superconducting Josephson junction. 

To conclude, we detected a Terahertz-range chaotic signal from a superconducting Josephson junction.
% and developed its proper theoretical description. 
 As this is the first such experimental observation of a chaotic signal emitted from a superconducting system in this frequency range, this finding presents a candidate for filling the Terahertz gap for chaotic oscillators.
The practical potential of TFFO chaotic generators are favored by the well established Nb technology.
It routinely used for fabrication of standard superconducting junctions which have established their reputation as reliable THz sources for spectral measurements both in the lab~\cite{Koshelets-spectrometers, Li-2012} and
in-field~\cite{Lange-2010, SIR-TELIS}.
%on board of high-altitude balloon for the atmosphere spectroscopy~\cite{Lange-2010, SIR-TELIS}.

Our results suggest that the range of possible applications of superconducting junctions are far beyond their present use: it should include devices for chaos-based communications and generators of random numbers 
%which are currently 
actively searched in various alternative technologies.
{
\drg
%Moreover, the found mechanism of generation of a THz chaotic signal
%Development of THz chaotic radiation sources 
The proposed mechanism for generation of chaos
opens horizons and opportunities of the developing noise-based molecular and biological spectroscopy~\cite{Koshelets-THz-spectroscopy}.  
The novel chaos-based THz spectroscopy can be used to probe physical and chemical processes in biological and living systems which otherwise would be impossible or extremely difficult to observe. 
%By taking instantaneous spectral snapshots with the use of a 
Broadband chaotic THz irradiation will enable to obtain a non-invasive spectral snapshot of an undisturbed system, which otherwise would be impossible to obtain by tools of the conventional spectroscopy.
In contrast to the conventional tools of spectroscopy, 
%the radiation sources with  
the chaos-based spectroscopy 
%the broad THz band 
may display a whole range of absorption lines simultaneously providing instantaneous radiation spectra in fast biological processes, transient biological structures, unstable molecules, radicals and chemical reactions. 

Remarkably, the TFFO stands out from all superconducting systems studied previously as a Josephson junction whose Josephson current oscillations are not at the Josephson frequency.

}

%The short interaction  time of non-invasive radiation  is  suitable  for  the  observation  of  
%This is particularly important for physical biological processes including  absorption, emission  and  detection  of  unstable sensitive biological transient structures, radicals and chemical reactions.
%This opens up a new research avenue for studying highly sensitive biological processes which were never studied so far. 

%-----------------------------------------------------------------------------------------------
%-----------------------------------------------------------------------------------------------
{\it Acknowledgments.}
%-----------------------------------------------------------------------------------------------
%-----------------------------------------------------------------------------------------------
The theoretical part of the work and numerical modeling are supported by the Russian Science Foundation under the grant 18-12-00429. The experimental study is supported from the grant no.~17-52-12051 of the Russian Foundation for Basic Research.
%The paper recognizes the use of the ``Hydra" High Performance System at Loughborough University. 


\begin{thebibliography}{99}
\bibitem{footnote1} {\drg 
Nb-AlN-NbN technology~\cite{NbN-Dmitriev, Nb-NbN-FFOs} allows to achieve higher radiation frequencies
as compared to Nb-$\rm AlO_x$-Nb junctions. The advantage of using Nb-AlN-NbN as opposed to NbN-AlN-NbN where both electrodes are made of NbN is achieving a better barrier quality. More specifically, while using the Nb-AlN-NbN junction we can achieve the normal to subgap resistance ratio $R_n/R_j$ to be about 30, the resistance ratio $R_n/R_j$ we could achieve for NbN-AlN-NbN contacts is limited by 10.}
\bibitem{Benettin-memory} For this, we extend dimensionality of the Lyapunov vector 
to hold information about the previous evolution. During the Benettin's successive renormalizations of  perturbation vectors 
along the given direction we also renormalize the memory variables.
\end{thebibliography}
\end{document}